# *In-situ* metallic coating of atom probe specimen for enhanced yield, performance, and increased field-of-view

Tim M. Schwarz[1*], Eric Woods[1], Mahander P. Singh[1], Xinren Chen[1], Chanwon Jung[1], Leonardo S. Aota[1], Kyuseon Jang[1,2], Mathias Krämer[1], Se-Ho Kim[1,3], Ingrid McCarroll[1], Baptiste Gault[1,4*]

1. Max-Planck-Institut für Eisenforschung, Max-Planck-Str. 1, Düsseldorf 40237, Germany
2. Department of Materials Science and Engineering, Korea Advanced Institute of Science and Technology, 291 Daehak-ro, Yuseong-gu, Daejeon 34141, Republic of Korea
3. *now at* Department of Materials Science and Engineering, Korea University, Seoul 02841, Republic of Korea
4. Department of Materials, Imperial College London, London, SW7 2AZ, UK

*Corresponding author E-mail address: tim.schwarz@mpie.de, b.gault@mpie.de

ORCID:

Tim M. Schwarz: https://orcid.org/0000-0001-9348-4160

Eric V. Woods: https://orcid.org/0000-0002-1169-893X

Mahander P. Singh: https://orcid.org/0000-0003-1784-7219

Xinren Chen: https://orcid.org/0000-0001-6528-8023

Chanwon Jung: https://orcid.org/0000-0002-9782-0261

Leonardo S. Aota: https://orcid.org/0000-0002-1520-2073

Kyuseon Jang: https://orcid.org/0000-0003-1826-7443

Mathias Krämer: https://orcid.org/0000-0002-1352-9064

Se-Ho Kim: https://orcid.org/0000-0003-1227-8897

Ingrid McCarroll: https://orcid.org/0000-0002-0584-4769

Baptiste Gault: https://orcid.org/0000-0002-4934-0458

# 1  Abstract

Atom probe tomography requires needle-shaped specimens with a diameter typically below 100 nm, making them both very fragile and reactive, and defects (notches at grain boundaries or precipitates) are known to affect the yield and data quality. The use of a conformal coating directly on the sharpened specimen has been proposed to increase yield and reduce background. However, to date, these coatings have been applied *ex-situ* and mostly are not uniform. Here, we report on the controlled focused ion beam *in-situ* deposition of a thin metal film on specimens immediately after specimen preparation. Different metallic targets e.g. Cr were attached to a micromanipulator via a conventional lift-out method and sputtered using Ga or Xe ions. We showcase the many advantages of coating specimens from metallic to non-metallic materials. We have identified an increase in data quality and yield, an improvement of the mass resolution, as well as an increase in the effective field-of-view. This wider field-of-view enables visualization of the entire original specimen, allowing to detect the complete surface oxide layer around the specimen. The ease of implementation of the approach makes it very attractive for generalizing its use across a very wide range of atom probe analyses.

# 2  Introduction

Atom probe tomography (APT) is a microscopy and microanalysis technique that provides three-dimensional compositional mapping (Larson et al., 2013; Lefebvre-Ulrikson et al., 2016; Gault et al., 2021). APT is based on the field evaporation of surface atoms (Forbes, 1995). A field in the range of 10–100 V/nm is generated by applying a high voltage on the order of a few kilovolts to a needle-shaped specimen, with an end radius in the range of 20–100 nm. Superimposed to this electrostatic field are either high-voltage pulses (Müller et al., 1968) or laser pulses (Kellogg & Tsong, 1980; Bunton et al., 2007) that enable time-control of the field evaporation and hence time-of-flight measurement for each ion. Conversely to most other microscopy techniques, the specimen also acts as the primary ion projection optics (Fortes, 1971; Cerezo et al., 1999; De Geuser & Gault, 2017), and control over its shape is hence critical to achieve optimal operation of the atom probe. Specimens can be prepared by electrochemical polishing (Melmed, 1991; Miller & Smith, 1989) or, as increasingly common, by lifting out a piece of the sample, depositing it onto a support and milling it into an appropriate shape using a dual beam scanning-electron microscope - focused-ion beam (SEM-FIB) (Prosa & Larson, 2017).

Needle-shaped substrates have been used as a support to deposit single or multiple layers of a range of materials and enable direct APT analysis. This approach to coat needles has been particularly useful to study materials that were otherwise difficult to prepare into sharp needles, including carbon-based materials for instance (Southon et al., 1975; Prosa et al., 2010; Zhang et al., 2021; Eder et al., 2017), a range of metallic multilayers to study diffusional and interfacial processes (Balogh et al., 2011; Larson et al., 2009; Tamion et al., 2006; Vovk et al., 2007), along with glasses (Greiwe et al., 2014) and thin organic layers (Kelly et al., 2009; Nishikawa & Kato, 1986) or even frozen liquids (Inghram & Gomer, 1955; Anway, 2003; Stintz & Panitz, 1991). The deposition of thin metallic layers on silicon (Jeske et al., 1999; Schmidt et al., 1988), ceramics (Seol et al., 2016) or glass (Kellogg, 1982) was shown to modify the field evaporation behavior, and for laser-pulsed atom probe tomography, it changes the light absorption and heat conduction properties of the specimen. These parameters are critical, since they control the rise and decay time of the "thermal pulse" that triggers field evaporation following laser pulse illumination (Houard et al., 2010, 2011; Vurpillot et al., 2009, 2006), and hence the achievable performance limits of atom probe microanalysis.

Coating of APT specimens has been reported following transfer from the SEM-FIB, through ambient atmosphere, to an external chamber by sputtering (Taylor et al., 2018), physical-vapor (Kim et al., 2022), chemical-vapor deposition (Felfer et al., 2014) (PVD & CVD, resp.) or atomic-layer deposition (ALD) (Li et al., 2021; Mosiman et al., 2021). Encapsulation of thin layers of liquid by using graphene conductive coating have also been proposed (Adineh et al., 2018; Exertier et al., 2021). These coatings were shown to lead to improvements in mass resolution and yield (Larson et al., 2013), with the material used and the thickness of the coating influencing the electric field itself (Kellogg, 1982; Adineh et al., 2017). The improvement in yield has been posited to be due to pores being filled (Barroo et al., 2020), as these are known to result in early specimen fracture, likely due to stress concentration (Moy et al., 2011; Wilkes et al., 1972).

Coatings on flat substrates have also been used to encapsulate nanomaterials (Larson et al., 2015), including nanoparticles (Felfer et al., 2015; Josten & Felfer, 2022), nanowires (Prosa et al., 2008; Lim et al., 2020) or nanosheets (Kim et al., 2020) prior to performing the preparation of atom probe specimens. Shadowing effects during deposition, or the chemistry of the deposited material, i.e. the Pt-containing precursor in the FIB (Perea et al., 2017), or the use of an electrochemical potential that can affect the material to be deposited (Kim et al., 2018), can however make the analysis and interpretation arduous.

Recently, Woods et al., 2023 revisited an approach, first introduced by Kölling & Vandervorst, 2009, to coat specimens directly in-situ in the SEM-FIB, at cryogenic temperature. This method used material sputtered by the ion beam from the surface of a piece of metal placed in the vicinity of the already sharpened specimen. Douglas et al., 2023 also showed how this in-situ deposition method could be used to strengthen the interface between a lifted-out sample and the Si-support, which proved critical to the yield for specimens prepared at cryogenic temperatures. This approach allows for precisely controlling which section of the specimen's surface is coated and presents the advantage of being versatile with respect to the deposited material and the thickness of the coating.

Here, we report a more systematic study on the influence of these coatings on the final atom probe results. We demonstrate coating using Cr, Ti, Al, In, Bi, Co and Ag on a Ga-FIB and/ or a Xe-plasma FIB. The coating materials were deposited on a range of metallic and non-metallic atom probe needles. We report on the composition and structure of the coating itself, and its multiple beneficial effects on the atom probe analysis, including increased yield, increased mass resolution, reduced background levels in laser pulsing mode, and an increased field of view. Ways forward with this approach are discussed for future possible generalization of the approach.

# 3  Methods

## 3.1  Materials

For this study, we selected a range of materials, which will be described in the relevant sections below. The targets used for in situ sputtering were typically small pieces of pure metals (99.99 %) either sourced from Goodfellow, Fisher Scientific GmbH or mixed leftover from the synthesis of larger ingots at the Max-Planck-Institute für Eisenforschung. The metal pieces were first cut into small 3x7x0.5 mm pieces and then ground and polished to create a flat surface and remove any oxidation products. Afterwards, the metal pieces were mounted on a stub for SEM or the copper clip from Cameca Instrument Inc.. The Fe-0.6 at. %B-0.2 at. %C thin films were prepared using physical vapor deposition (BesTeck). The process involved co-sputtering from a pure iron target (99.995%, Mateck, Germany) and a boron target (99.9%, Kurt J. Lesker, USA) onto a MgO substrate. After sputter deposition, the thin films were annealed at 300 °C for 40 hours and then cooled to room temperature inside the vacuum chamber.

## 3.2  FIB-based coating

To highlight the versatility of the approach, multiple SEM-FIBs were used to prepare specimens, a Helios 600i, a Helios 5 CX Ga SEM-FIB, and a Xe-plasma FIB-SEM Helios PFIB (all Thermo-Fischer Scientific, Hillsboro, OR, USA). The latter two instruments are equipped with EZLift tungsten micromanipulators, with the Helios 5 manipulator having cryogenic capabilities, whereas the 600i is equipped with an Omniprobe micromanipulator.

**Lamella preparation**

(A) Step 1   (B) Step 2   (C) Step 3

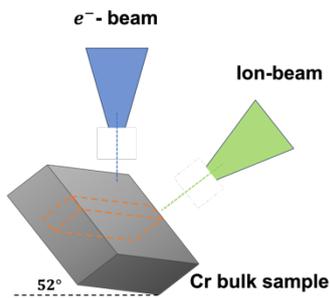 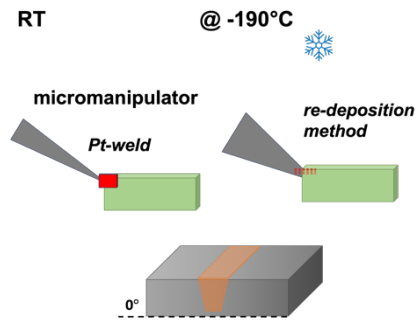 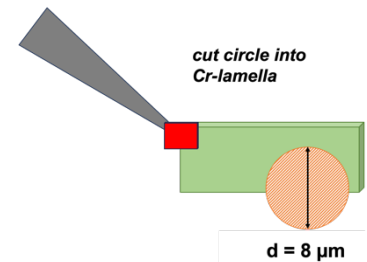

Figure 1: Schematic view of the target preparation. (A) the trenches are cut at 52° and then (B) the lamella is attached to the micromanipulator at 0°, either at with the GIS or with the redeposition, which is preferred at cryogenic temperature. (C) The opening with a diameter of 8 µm is then cut into the lamella.

Figure 1 summarizes the lift-out procedure as described by Thompson et al., 2007: a lamella of the metal to be deposited, with typical size of 20x10x2 µm, is lifted out and attached to the micromanipulator (Step 1, Figure 1A). This can be done either by welding using the gas injection systems (GIS) at room temperature or by re-deposition under cryogenic temperatures (Schreiber et al., 2018; Woods et al., 2023) (Step 2, Figure 1B). A semi-circular opening with an 8–10 µm diameter is then cut into the lamella (Step 3, Figure 1C).

**Coating of APT specimen**

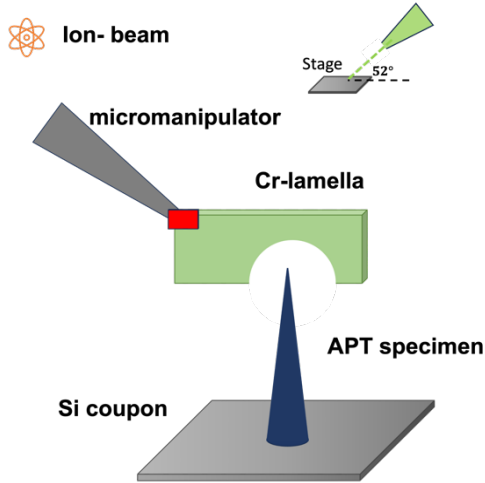
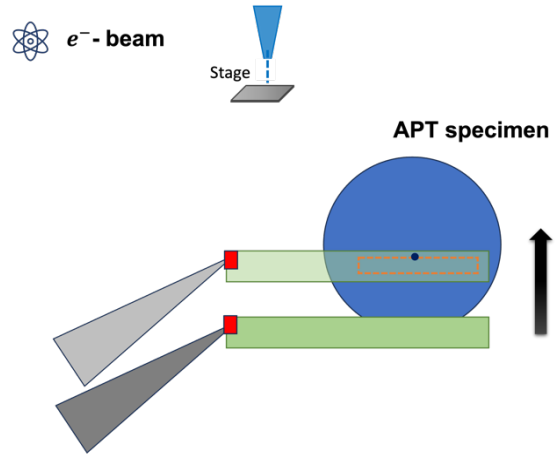
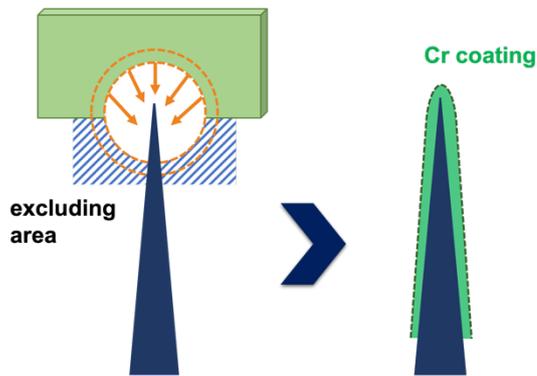
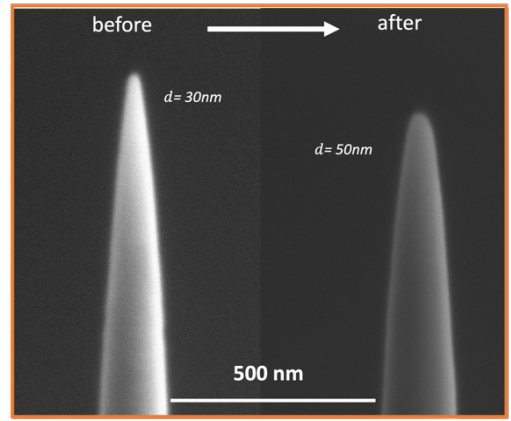

Figure 2: Schematic of the different steps for *in-situ* coating of specimens in the FIB. (A) Show the positioning of the lamella in the ion beam in such a way that the specimen is in the center of the opening. (B) The alignment in the electron beam is shown, where the top view of the specimen is represented by the blue circle, while the arrow indicates the movement of the manipulator to bring the target to the desired position. (C) The ion beam is screened in the orange areas, while the blue area represents the exclusion zone. (D) SEM images before and after coating a Ni tip with Cr are shown.

The steps for the coating of the needle-shape specimens are detailed in Figure 2. In the first step, in Figure 2A, the opening in the lamella is placed over a sharpened APT specimen, so that it sits directly in the center of the opening as imaged by the ion beam (Step 1). The distance between the bottom edge of the circle is typically about 1.5 µm from the specimen's tip. The distance from the lamella to the tip and the diameter of the circle will determine the length of the specimen is coated. In the next step, in Figure 2B, the lamella is placed in the electron beam in such a way that the sharpened specimen is covered by the upper edge of the lamella (Step 2). Finally, the ion beam is rastered on the metal target, from the inner circle outwards, using the pattern with the diameter of 12/8 µm and 1 µm in z- direction shown in orange in Figure 2C (Step 3). To ensure that the ion beam does not hit and damage the specimen itself, an exclusion zone is added that covers the specimen, as indicated by the blue shaded area in Figure 2C.

Following a first deposition, the specimen stage is rotated by 90°, and a new layer is deposited. This process was repeated until a full 360° rotation was performed. Each side of the specimen was hence coated twice, which appeared to provide a sufficiently homogeneous and uniform coating for smooth atom probe analysis. Figure 2D shows an example of a Ni specimen, before and after coating with Cr. On this scale, the coating appears in the form of a homogeneous layer and indicated by a uniform increase in diameter of the specimen.

It is important to keep in mind that the parameters used depend on the material being sputtered and the desired film thickness. For Cr, sputtering was performed at 30 kV and 40 pA for 30 seconds. Note that the pattern settings are based on Thermofisher calculations for Si and are given in Table S 1. The parameters may vary using different instruments. The average time required to coat six samples from four sides is approx. 1 hour, excluding preparation of the lamella.

In addition to coating needle-shaped samples, it is also possible to coat the surfaces of flat samples. Protection layers deposited using the Pt-precursor from the GIS are generally not perfectly dense and contain carbon residues from the precursor gas, which can then cause issues during imaging by transmission-electron microscopy, and often make field evaporation in the APT of this layer problematic. When deposited with the ion beam, there can also be structural damage in the sub-surface region (Thompson et al., 2007). With the method described in Figure 3, a metallic layer can be deposited *in-situ* on the desired location of the sample's surface. The lamella is moved close to the surface and the ion beam is rastered over the rectangular area marked in orange, Figure 3A, B.

A higher-magnification SEM image shows a clear contrast between the deposited Cr layer and the underlying substrate, Figure 3C. The layer thickness is strongly dependent on the sputtering time. The lamella remains static in one position during sputtering as the coated area is significantly larger than the selected pattern due to the beam scattering. If a larger region of interest (ROI) is required, the lamella can be moved, and the coating process repeated. The same parameters were used as for the needle-shaped specimen, but a longer deposition time of 1 - 4 min was required. In both cases, the coating can be done under cryo-conditions, making this method very versatile. Using the GIS under cryo-conditions is problematic because the gas condenses on all cold surfaces, which makes a controlled deposition impossible. Using this method, a thin metal layer can be deposited under cryo-conditions in a controlled manner.

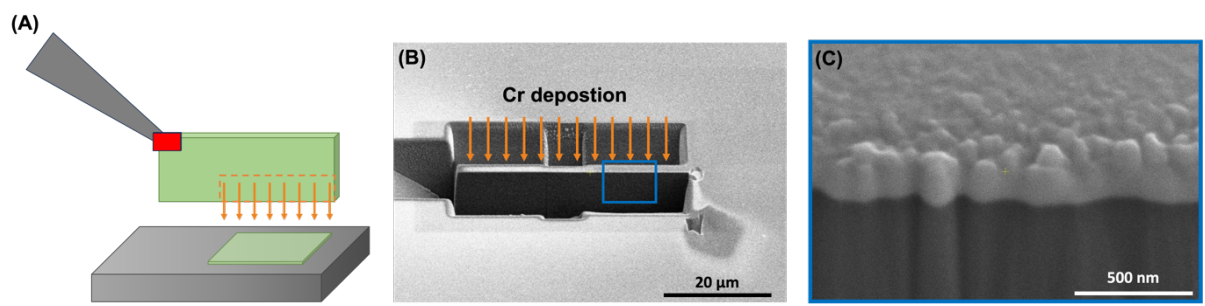

Figure 3:(A) schematic overview of the coating of a flat surface and in (B) the SEM image showing the coated surface. In (C), an SEM image with higher magnification the contrast between Cr and the substrate is clearly visible.

## 3.3 Atom probe tomography

Following preparation and coating, specimens were either transferred through air or via cryo/vacuum suitcase (Ferrovac) to a Cameca LEAP 5000XS (straight flight path) or 5000XR (reflectron) atom probe. The specific running conditions will be specified below for each set of results. Data reconstruction and analysis were performed using APSuite 6.3 software. The initial reconstruction was performed using the measured start radius and shank angle of the specimen obtained from the SEM image of the sharpened/coated specimen, following the point projection protocol by Geiser et al., 2009. Where possible, the reconstruction parameters, image compression factor (ICF) and field factor ($k_f$) were calibrated using crystallographic poles as described in Gault et al., 2009 and adjusted to obtain lattice spacing reported in the literature.

## 3.4 (S)TEM

For the transmission electron microscope (TEM) analysis, a FEI Titan Themis microscope with an image corrector operated at 300 kV was used to reveal the structure of the *in-situ* metallic coating of a Ni specimen. Energy-dispersive X-ray (EDX) analyses and high- angle annular dark field (HAADF) imaging was performed in scanning TEM (STEM) mode using the same instrument. For the 4D (S)TEM of the Fe(BC) specimen, a JEOL2200FS transmission electron microscope (TEM) was operated at 200 kV. EDX analyses and high-angle annular dark field (HAADF) imaging were performed in scanning TEM mode using the same instrument. Scanning Precession Electron Diffraction (SPED) (Harrison et al., 2022) was executed on the same TEM, with a measured beam size of ~1 nm and a 0.82º precession angle supplied by a Digistar hardware unit (NanoMEGAS SPRL). Each SPED dataset covered a scan frame comprising (255 × 276) pixels at a frame rate of 24 frames per second. Since the electron probe intensity distribution was strongly peaked, the datasets were indexed using an Automated Crystallographic Orientation Mapping (ACOM) tool.

## 4 Results & discussions

### 4.1 Structure and composition of the coating

Specimens for APT were prepared using a Ga-FIB and a Xe-PFIB from nano-crystalline, high purity (99.99 %) Ni. Through grinding of a Ni plate, which had been strongly deformed resulting to a nanocrystalline structure, specimens were prepared by FIB lift out, as outlined in Thompson et al., 2007. In addition, specimens were prepared from the B-doped Fe film (see methods above). TEM and (S)TEM analyses were performed to study the structure and thickness of the deposited Cr layer deposited on pure Ni (fcc) and Fe (bcc).

Figure 4A displays TEM images after performing the sputter coating only once. A thin homogeneous layer has been deposited on the Ni specimen. After coating the specimen in each pass and rotating it 90° four times to coat it from all sides, the deposited layer thickness is approx. 10 nm. This is consistent with our estimation form the SEM images, shown in (Figure 2). Interestingly, the Cr layer is conformal with the specimen, homogeneous and uniform over several hundred nanometers down the shank, as readily visible in Figure 4B. There seems to be no discernible shadowing effects or columnar growth of the sputtered layer, which have been previously reported in the literature (Prosa & Larson, 2017). EDX mapping in S(TEM) mode shows the Cr layer around the Ni needle (Figure 4C).

Figure 4D is a close-up of a notch on the edge of the specimen. Following coating, the outer edge of the specimen is smooth and homogeneous, which is likely to reduce the stress localization, leading to increased mechanical stability of the specimen during analysis. However, it should be noted that the evaporation fields of the sample and the coating must be adjusted to avoid inhomogeneous evaporation (Larson et al., 2013), which leads to locally stronger fields and thus possible higher stress between the coating and the sample. From the (S)TEM images, the sputtered Cr layer appears amorphous on the Ni specimen (Figure S1).

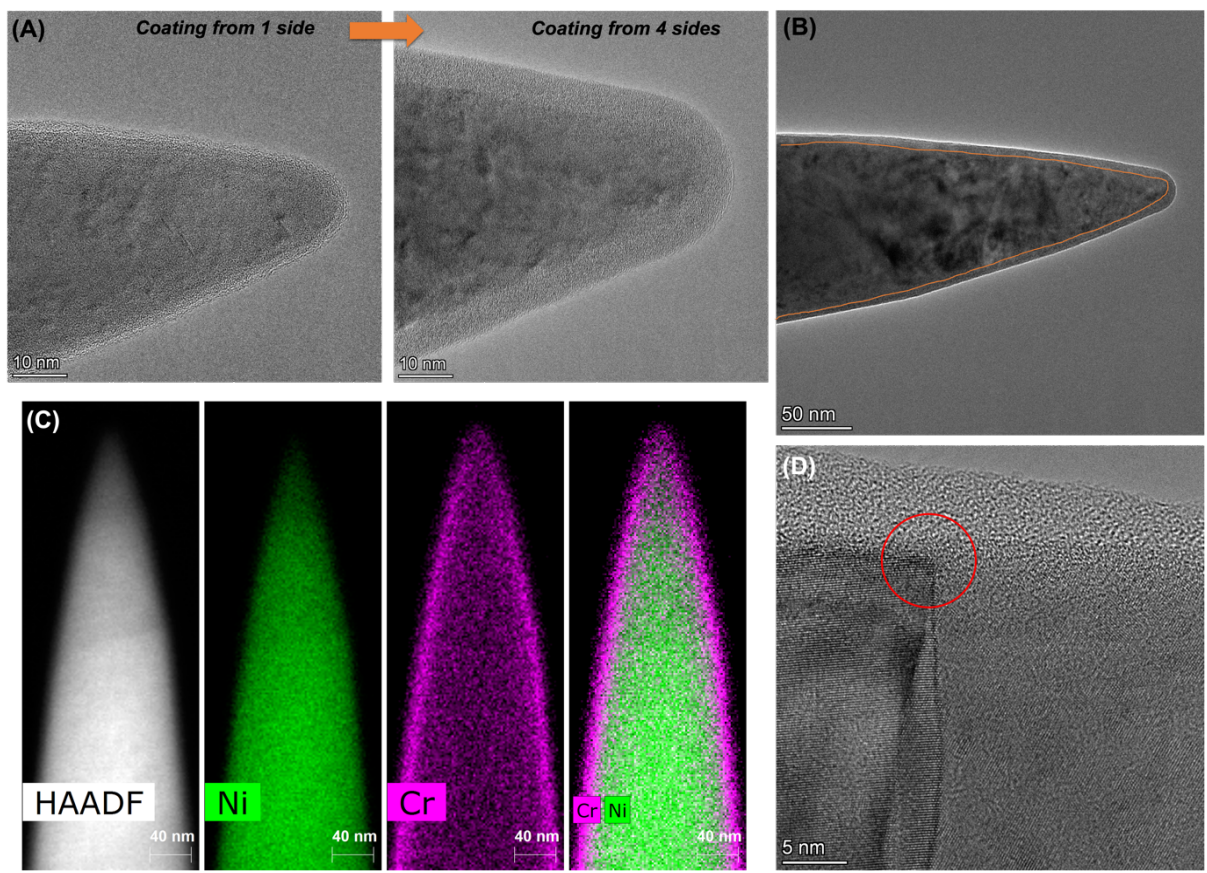

Figure 4: (A) Shows a Cr coating on a Ni specimen from one side and from four sides for comparison. (B) Shows the tip with a homogeneous Cr layer of several hundred nanometers along the shaft of the tip with a coating thickness of approx. 10 nm. In (C) the EDX mapping shows the Cr coating around the Ni specimen. (D) Defects and roughness of the sample surface can be homogeneously coated, as shown here on the example of a grain boundary.

Figure 5A shows the HAADF-STEM images of the coated Fe-specimen from four sides. Figure 5B, C, are PED patterns pertaining to the crystal structure of Cr and obtained from different parts of the coating. ACOM-SPED indicates that the coating is nanocrystalline in part, with regions that could not be indexed and could be considered amorphous, Figure 5D. Due to the very close lattice constants of α-Cr and α-Fe, the indexing is not unambiguous and therefore has some inaccuracies. 4D-STEM data also collected from the same specimen allows to distinguish α-Cr and α-Fe through the specimen.

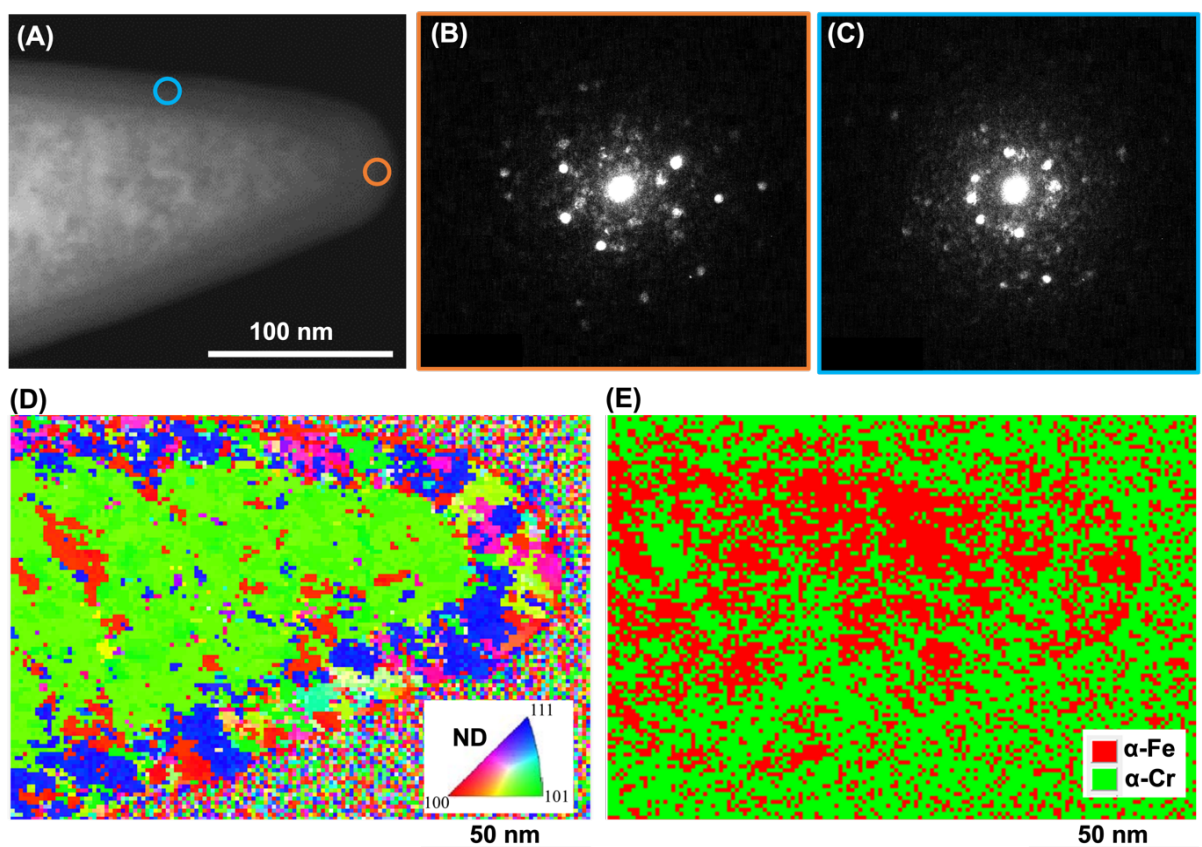

Figure 5: (A) HAADF-STEM image of the Fe-1%B specimen after coating. (B) and (C) show the PED pattern from the orange and blue circle area and (D) visualizes the indexing of the 4D-STEM dataset. E shows the phase detection of 4D-STEM data sets.

The results in Figures 4 and 5 suggest that the structure of the coated layer depends on the underlying substrate. If the lattice structures of the coating and the substrate is too different, an amorphous layer will first form to compensate for the lattice mismatch between the two crystal structures. As the coating grows thick, and depending on the mismatch between the two crystal structures, the coating can become crystalline.

Studies have shown that Cr grows amorphously on a glass substrate and has a crystalline structure on a crystalline substrate, supporting our observations on different substrate materials (Khanna et al., 2006; Miller & Holland, 1997). This can explain the amorphous Cr-layer seen on Ni, and the nanocrystalline coating on Fe, on relatively thin coatings (10-20 nm).

From the perspective of the APT measurement, an amorphous layer may be preferable as they would leads to little or no stress on the sample's surface (Larson et al., 2013), whereas nanocrystalline layers can exhibit heterogeneous field evaporation at defects of for different sets of atomic planes. From a heat conduction perspective, the opposite could be expected. It is possible that the structure depends not only on the substrate material, but also on other parameters such as the use of Ga or Xe ions, ion beam acceleration voltage and current. The deposition under cryogenic temperatures could also favor the amorphous formation of the Cr coating. All these parameters will be explored in future.

The chemical composition of the Cr-coating depending on the use of an Xe and Ga FIB on a Ni specimen was then analyzed by APT following transfer through ambient conditions to the atom probe. Figure 6A and 6B display a 10 nm slice of the reconstructions of specimens prepared by Xe-PFIB and Ga-FIB, respectively, at a chamber pressure of aprrox. $2.05 \cdot 10^{-6}$ mbar. The coating covers the top and sides of the specimen. The Xe and Ga distributions are visualized using iso-surfaces. Xe tends to agglomerate, whereas Ga appears distributed across the entire Cr layer. The composition of the Cr layer varies strongly, with 55.01±1.51 at. % Cr, 44.31±1.59 at. % O and 0.68±0.09 at. % Xe obtained on the PFIB (Table 1). This contrasts with the composition of the coating obtained by Ga-FIB that is 87.19±1.40 at. % Cr, 7.36±0.99 at. % O and 6.02±1.46 at. % Ga (Table 1). Deposition of the Cr layer under cryogenic conditions on a Ga-FIB shows that the Ga implantation in the deposited layer is significantly lower by a factor of 3.12. The composition of the coating obtained in cryogenic condition is Ga-FIB that is 90.60±0.66 at. % Cr, 7.47±0.87 at. % O and 1.93±0.21 at. % Ga (Table 1). In contrast, the oxygen concentration in the Cr layer does not change much, while the pressure in the FIB is in the range of approx.. $8.05^{-7}$ mbar.

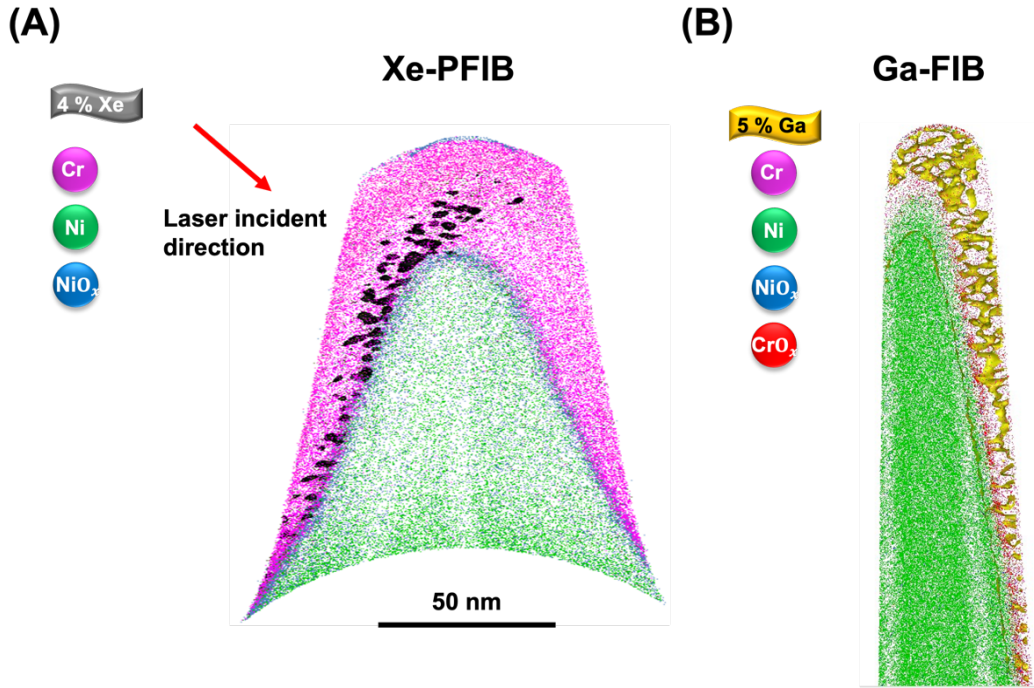

Figure 6: Comparison of the structure of the Cr layer produced with a Xe and Ga FIB. In (A) the Xe clusters with an iso-surface concentration of 4 at.% and in (B) the Ga distribution with an iso-surface concentration of 5 at.% are visualised.

Table 1: Average composition of the Cr coating prepared by a Xe-PFIB (3 specimens) and Ga-FIB at room temperature (5 specimens) and cryogenic temperature (3 specimens). All numbers are given in at.%.

|    | Xe-PFIB     | Ga-FIB      | cryo Ga-FIB |
|----|-------------|-------------|-------------|
| Ga | -           | 6.02±1.46   | 1.93±0.21   |
| Xe | 0.68±0.09   | -           | -           |
| Cr | 55.01±1.51  | 87.19±1.40  | 90.60±0.66  |
| O  | 44.31±1.59  | 7.36±0.99   | 7.47±0.87   |

Oxygen likely originates from the fast formation of the passivating Cr-oxide in between two passes of the ion beam on the surface of the Cr lamella. The difference in composition could result from the fact that the Xe removes ions mostly from the surface, which is likely to be more enriched in oxygen. In general, a higher Ga content is observed compared to Xe, which can be attributed to its higher reactivity, deeper penetration, and overall higher concentration of Ga inside of the target material compared to Xe.

The use of Ga-FIB under cryogenic conditions significantly reduces the implantation of Ga, but it appears to have no significant effect on the oxygen concentration in the film from by sputtering from the cryogenically cooled Cr-lamella. The oxygen concentration thus depends significantly on the formation of the oxide layer on the surface of the lamella, or at least the adsorption of O-containing species which can be facilitated at low temperature. These issues could be prevented by using an ultra-high-vacuum (UHV) FIB.

An uneven distribution of Xe can be seen in the reconstruction (Figure 6A). It has been shown in the TEM by Wittmaack & Oppolzer, 2011 and in the APT studies by Estivill et al., 2016 of that Xe forms clusters in the substrate material during specimen preparation. The same phenomenon can be observed in the coated Cr layer by Xe ions. The non-uniform distribution of Xe clusters could be due to the position of the laser beam hitting the specimen (Figure 6A). The relatively higher increase in the specimen temperature on that side, leading to slightly lower field enabling their field evaporation, whereas on the opposite side, preferential evaporation of Xe at the standing electrostatic field is more likely to occur.

## 4.2 Yield and mechanical resistance

Atom probe specimen from the same nanocrystalline Ni were prepared and Cr-coated using a Ga-FIB (Helios 5 CX). APT analyses were performed on a LEAP 5000 XS at 50 K, in laser pulsing mode (laser pulse energy of 40 pJ), with a repetition rate of 200 kHz and with a target detection rate of 1.5 ion per 100 pulses, on average, for both uncoated and Cr-coated specimens. Four uncoated Ni specimens were analyzed. The voltage curve from one is plotted in Figure 7A. It exhibits multiple dips, marked with blue arrows in Figure 7A, that are related to grain boundaries as proven by detector hit maps, shown in Figure S2 A-C. These dips do not correlate with any detectable changes in the evaporation rate (Figure S2D), or specimen fracture (Gault et al., 2012). In each case the specimen fractured before 6 million ions could be collected. After coating the nanocrystalline Ni specimen, the voltage dips are no longer observed, as can be seen from the voltage curve in Figure 7b. For three analyzed Cr-coated specimens, the analysis was stopped when a set voltage of 8 kV was reached, and for each data set 60–100 million ions were collected without specimen fracture occurring.

We attribute this increase in yield and sample throughput to more stable field evaporation conditions. The high density of grain boundaries in nanocrystalline materials makes them particularly susceptible to fracture. The Cr coating appears to smooth the surface roughness out, and fill in some of the notches or pores, as observed by TEM in Figure 4D, thereby explaining the improved field evaporation conditions. Mechanical stabilization and the associated improvement in yield and throughput of coated samples was observed for the measurement of nanocrystalline Ni here and for a range of different material systems (Singh et al., 2023; Woods et al., 2023), which is in agreement with previous work (Larson et al., 2013; Kölling & Vandervorst, 2009).

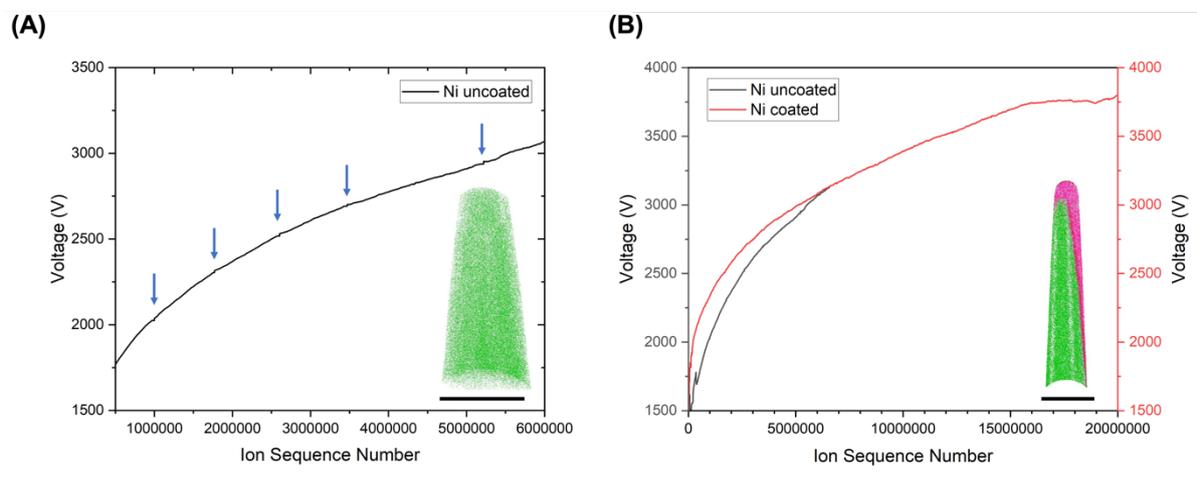

Figure 7: In (A), the voltage curve for the uncoated Ni sample shows several dips, indicating structural changes during the measurement due to GB or other defects and inhomogeneities. (B) Compares the voltage curve for the uncoated and coated Ni samples, showing a monotonous increase for the coated sample with no noticeable dips.

## 4.3 Mass spectrum improvements

Improvement in the mass resolution and background levels with a thin metallic coating of poor thermally conducting specimen have previously been reported (Seol et al., 2016; Larson et al., 2013). Here, a small fragment from the dentine of a tooth of an unidentified dinosaur, found in South Korea, was analyzed by APT on a LEAP 5000 XR. APT analyses were performed at 50 K, in laser pulsing mode (laser pulse energy of 60 pJ), with a repetition rate of 100 kHz and with a target detection rate of 1.0 ion per 100 pulses, on average, for both an uncoated specimen and a Cr-coated specimen. Specimens were prepared on a Ga-FIB (Helios 600i) using the protocol in Thompson et al., 2007.

Biominerals, such as teeth, consist mainly of hydroxyapatite where Ca is the main constituent, along with other inorganic elements (Mg, Zn, P, …). They typically have poor thermal diffusivity, and therefore tend to form more pronounced thermal tails during APT analysis (Gordon & Joester, 2011; Gordon et al., 2012). In addition, these materials have poor absorption properties for the laser, so that a higher laser pulse energy is required to obtain field evaporation under electrostatic field condition that do not lead to appreciable background levels. The tails of the mass peaks lead to a loss of information, particularly for impurities and trace elements that may be indistinguishable below the thermal tails.

To investigate the influence of the Cr coating on improving thermal conductivity, some dentine samples were coated with Cr and others were not. Figure 8 shows a section of the mass spectrum obtained from the analysis of a non-coated sample (blue) with over 11 million ions, and a coated sample (red) with over 21 million ions. Both spectra have been normalized to the $^{40}Ca^{2+}$-peak at 20 Da.

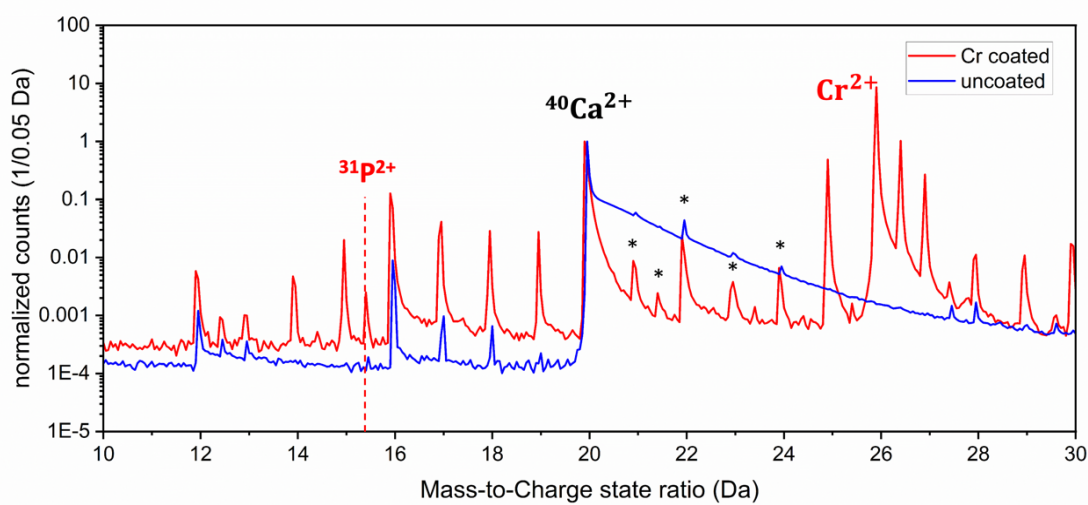

Figure 8: Normalized mass spectra to the Ca-peak for a coated (red) and uncoated (blue) specimen prepared from the dentine of a fragment of an unidentified dinosaur tooth. An asterisk indicates the different isotopes of Ca.

In the uncoated samples, the long thermal tail of the $^{40}Ca^{2+}$ peak leads to a considerable overlap with the peaks of the other Ca isotopes. The thermal tailing spreads over a long area, indicating poor thermal conductivity of the material. Such long tailings can make the identification and quantification of trace elements impossible. It is also difficult to identify isotope ratios, which can be critical in quantifying data when there are overlapping peaks and for the dating of fossils and minerals (Reddy et al., 2020). The shape of the Ca peak is typical for (bio)minerals obtained from a different biominerals in the literature (Grandfield et al., 2022).

In the case of the Cr-coated specimen, the mass resolution is vastly improved by reducing the thermal tailing, which greatly simplified the peak identification and separation between signals. This facilitates the detection or quantification of peaks in the thermal tailing, e.g. $^{31}P^{2+}$ at 15.5 Da, which were not possible to detect without Cr coating. In addition a more accurate isotopic ratio measurements are possible: for the uncoated specimen the $^{40}Ca^{2+}$ peak represents only less than 72 at. %, whereas it is 94.9 at. % for the coated specimen, noting that the natural abundance is 96.9 at. %. With the Cr coating, it is also possible to detect other low abundance isotopes of Ca, whereby it should be mentioned that the Peak at 23 Da has an overlap with $^{23}Na^+$ and is therefore higher.

We did not perform specific quantification of the improvement in mass resolution or background level since the specimen geometry (radius, shank angle, length) varies from specimen to specimen, even over the course of a single analysis. The change in shank angle alone affects substantially the mass resolution of the atom probe data, as demonstrated experimentally and computationally by Perea et al., 2008.

## 4.4 Effect on the field-of-view: analysis of aluminum

### 4.4.1 Pure-Al analyses

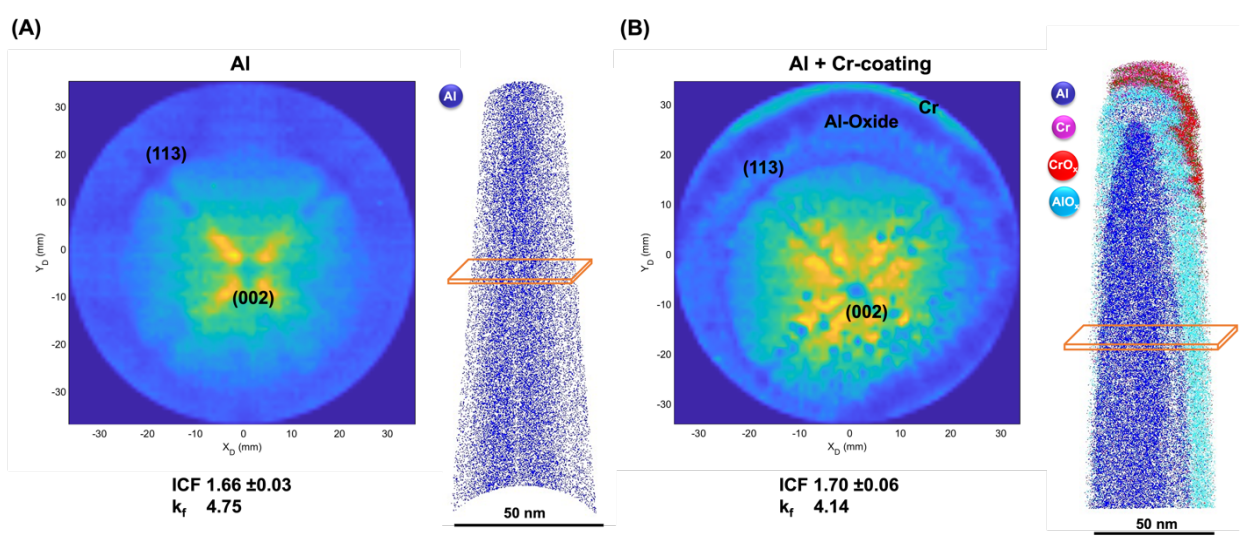

Figure 9: (A) data for the uncoated Al specimen clearly shows the (002) and (113) poles in the detector hit maps, along with the 3D reconstruction. (B) The Cr coated Al specimen shows the same poles in the Al-core, and the passivating surface oxides

*around the specimen indicates a larger field of view. The extracted area for the desorption map is marked with an orange parallelepiped.*

Next, we wanted to investigate the effect on the field-of-view and possible changes in the image compression factor (ICF). Specimens were prepared from a pure Al plate, ground, and polished to remove surface oxides, using the Xe-PFIB to minimize any spurious effect of Ga implantation on the visibility of crystallographic poles. To accurately determine the change in image compression factor and field-of-view, while avoiding the influence of different radii or shank angles from specimen to specimen, we first measured 8 million ions from a specimen, then performed the Cr-coating and measured the same specimen again. The APT analyses were performed on a LEAP 5000 XS at 50 K, in laser pulsing mode (laser pulse energy of 40 pJ), with a repetition rate of 200 kHz and a target detection rate of 1.5 ions per 100 pulses, on average. The same parameters were used for the uncoated and Cr-coated samples to aim for consistent measurement conditions. Figure 9A and 9B report on data obtained on the pure Al specimen before and after coating respectively. Since the specimen transfer from the atom probe into the FIB was done in ambient atmospheric condition, a thin oxide layer has formed, as visible in Figure 9B. Similar results were obtained for three uncoated, and later coated, specimens.

To study a native oxide on an APT sample it is necessary to capture the first thousands of atoms from the tip apex. This can be very challenging due to alignment requirements and due to their tendency to fracture off prior to reaching the metal/oxide interface. Here, after Cr-coating, the native oxide layer on the tip apex and along the shank of the specimen can be analyzed, leading to a massive increase in square nanometers of analyzable oxide/metal interface data. The capability of studying the surface of the atom probe tip is significant for any research group studying surface catalytic or oxidation and could lead to a higher success rate and throughput in the future. The thickness of the natural oxide layer on Al was measured based on the APT analysis to be between 6-8 nm, which is in good agreement with literature values (Evertsson et al., 2015).

### 4.4.2 Effect on the reconstruction parameters

Detector hit maps from the analysis of the specimen prior to coating, and after the coating show clear crystallographic poles, Figure 9A and 9B, respectively. The ICF depends on the theoretical and observed crystallographic angles between two poles (Gault et al., 2008). Poles pertaining to the (002) and the four variants of the (113) planes were used to calculate the ICF, which is 1.66±0.03 and a calculated $k_f$ of 4.75 following calibration. After coating the same specimen with Cr, Figure 8B, the ICF was calculated using the same poles when visible, and it was 1.70±0.06, with a $k_f$ of 4.14 determined from adjusting the (002) plane spacing at a depth of approx. 150 nm into the Al.

The field evaporation of the surface atoms results in a progressive increase in the tip's radius that typically results in a decrease in $k_f$ and ICF (Gault et al., 2011; Loi et al., 2013). Here, the decrease in $k_f$ can be explained by the increase in the specimen's radius due to the coating (Gault et al., 2011; Loi et al., 2013). However, contrary to the expectation that the ICF should drop with increasing radius (Gault et al., 2011; Loi et al., 2013), we see no significant change in the ICF after the Cr coating has been evaporated from the top and the Al core is being analyzed – maybe a slight additional compression of the field lines and therefore of the ion trajectories arising from the presence of the Cr coating, but no severe distortions towards the interface between the metal and the coating.

### 4.4.3 Origins of the increased field-of-view

A striking difference following coating is the analysis of the native oxide on the specimen's shank, which indicates a significant increase in the imaged surface area at the specimen, despite a limited change in angular compression, i.e. the ICF is almost constant, as illustrated in Figure 10. There are two important factors here. First is the angular field-of-view, that is typically between ±25–30° about the specimen's main axis, and that relates to the angular compression of the trajectory reflected by the ICF (De Geuser & Gault, 2017). Second, this angular field-of-view defines the actual field-of-view that is the imaged area at the specimen's surface shown by the red arc in Figure 10A, and that corresponds to the reverse projection of the detector onto the emitting specimen. As the radius increases because of the field evaporation during the APT analysis, the imaged surface area also increases, leading to a total analyzed volume delineated by the thick red lines.

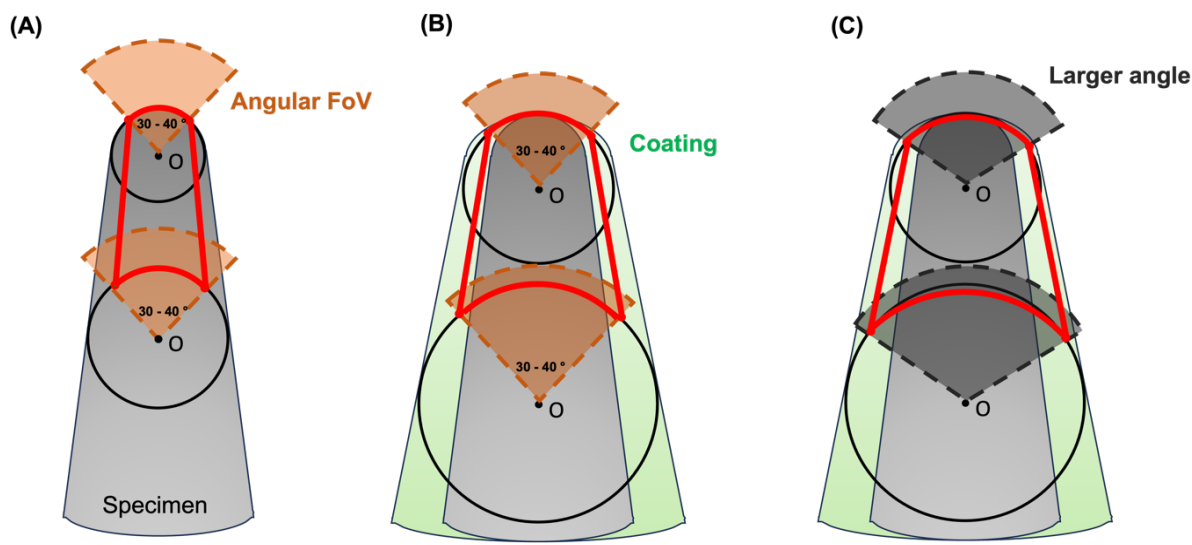

Figure 10: (A) Schematic view of the angular field-of-view of an uncoated sample, in red is delineated the analyzed volume by APT. (B) The coating makes the specimen thicker, thereby increasing the imaged surface area at the specimen's tip, even with a constant angular compression. (C) A larger angular field-of-view arising from a change in the ICF would simply make the field-of-view mildly wider.

Coating the specimen is equivalent to increasing its radius. Even if the angular field-of-view (i.e. ICF) remains the same, as in Figure 10B, the imaged surface area increases, allowing for imaging the sides of the original specimen. A similar increase in the imaged surface could arise from an increased angular field-of-view, Figure 10C, yet this would have translated into a strong change in ICF, which was not measured experimentally here, but had been reported previously (Du et al., 2013). Possible changes in the thickness of the coated film would primarily lead to changes in the specimen's end radius, leading to changes in the voltage to reach the electrostatic field necessary to reach the set detection rate, and not much on the compression of the ion trajectories, as suggested by Figure 10B. Measurable changes in the voltage were not observed experimentally, suggesting a rather conformal and homogenous coating. The application of this thin coating appears to lead to an increased field-of-view similar to the use of an electrostatic lensing system (Bostel & Yavor, 2010; Tegg et al., 2023). The use of the conventional reconstruction algorithm did not lead to severe distortions on the edges of the field-of-view.

## 4.5 Effect on the field-of-view: precipitates in Al-alloy

To demonstrate the influence of the change in the FoV, a similar set of analyses on an Al-5.18Mg-6.79Li (at. %) alloy aged for 8h at 150°C to form spherical δ'-precipitates, for more details, please refer to Gault et al., 2012. Following grinding and polishing to remove surface oxides, specimens were prepared using the Xe-PFIB. APT analyses were performed on a LEAP 5000 XS at 50 K, in laser pulsing mode (laser pulse energy of 20 pJ), with a repetition rate of 200 kHz and a target detection rate of 1 ion per 100 pulses, on average, for both the uncoated and Cr-coated samples.

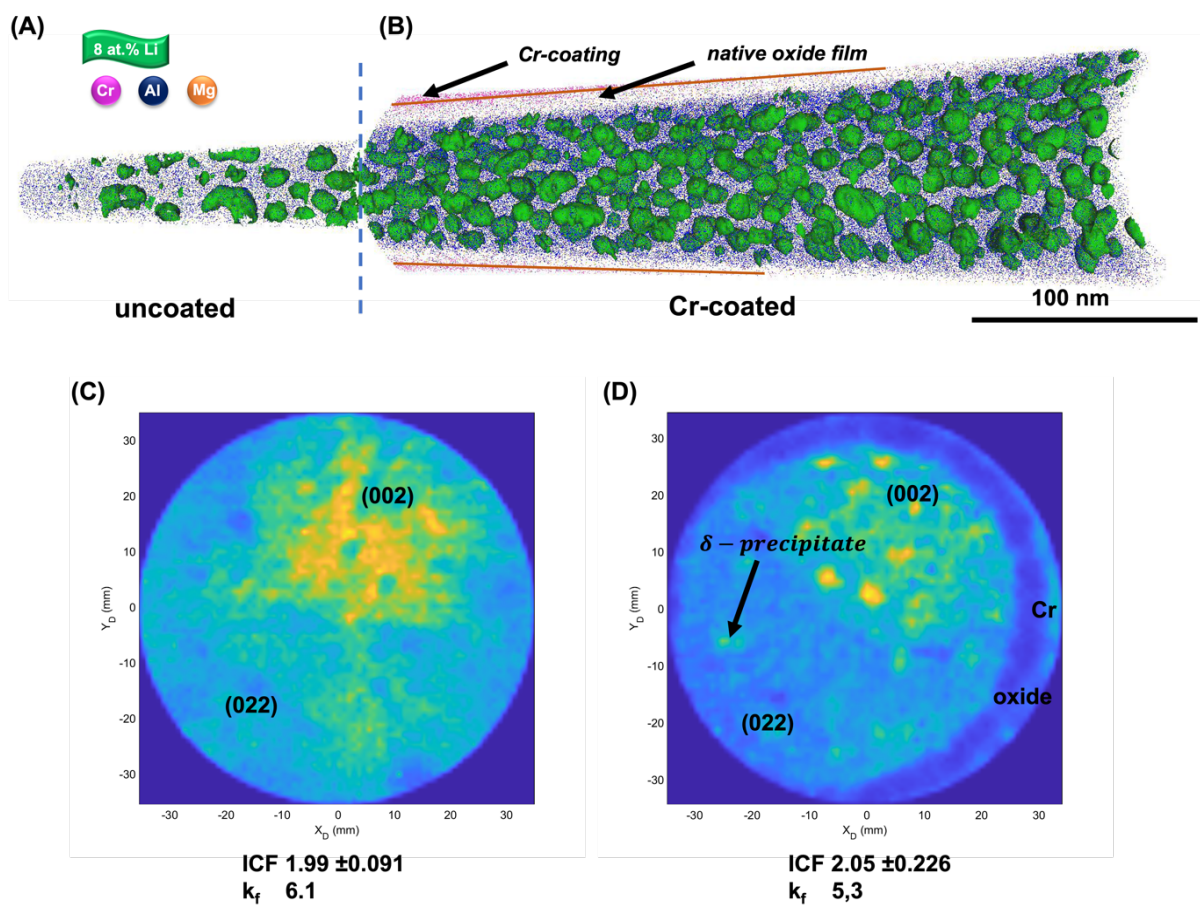

Figure 11: In (A) a comparison between uncoated and coated samples, Li δ'-precipitates are visualized using an iso-surface concentration of 8 at. % Li. In the Cr-coated sample (B), the Cr layer and the naturally formed oxide layer can be seen, the interface between the two being highlighted by orange lines. (C) and (D) show the corresponding desorption maps for the coated and uncoated specimen.

Figure 11 shows a 10nm thick slice through the combined reconstructions obtained from the analysis of the same specimen uncoated (Figure 11A) and Cr-coated (Figure 11B). The ICF was determined using the two poles marked in Figure 11C and D, and the $k_f$ factor was calculated by adjusting the spacing of the (002) planes. Like the pure Al specimen discussed above, no significant change in the ICF is observed. The increase in field-of-view offered by the Cr-coating allows for imaging the native oxide and part of the Cr layer itself, with the Cr/native oxide interface indicated by orange lines. It also leads to a substantial increase in the number of precipitates imaged through the specimen's cross section. This increase in the number of analyzed precipitates for a single specimen in one measurement can help improve statistical accuracy, and it is also possible to analyze larger precipitates that may not be directly in the center of the sample. These were also highlighted as advantages arising from the use of the lensing system in other instrument design (Bostel & Yavor, 2010; Tegg et al., 2023), but the same effect can be achieved by the thin metallic coating of the specimen using a conventional atom probe geometry.

## 4.6 Discussion

### 4.6.1 Perspective on the benefits of coatings

Several effects of an *in-situ* deposited metallic coating on APT specimens were reported, including improved mechanical resistance, better thermal conductivity, and an increase in the field-of-view. It is interesting to note that the ICF does not change significantly, indicating only a small additional compression of the field lines. A major advantage of this method is that no additional instrumentation is required, and the coating can be deposited in any FIB, with a Ga or Xe source and under cryogenic conditions.

In addition, the coating of the samples allows the complete oxide layer around the tip to be seen, as well as the Cr layer, thus allowing the outer layer and the thickness of the oxide layer to be analyzed. In the recently released Invizo instrument, an electrostatic lensing-system compresses the trajectory of the ions to increase the FoV up to 90-100%. Tegg et al., 2023 reported the analysis of the outer oxide layer formed on a Zr specimens, however, as full cross section of the specimen cannot be imaged, the outermost part of the oxide in contact with the environment is not analyzed over the full length of the specimen, hindering the analysis of the oxide layer through it full thickness. To confirm that the entire specimen's cross-section was analyzed often requires the use of microstructural features at the periphery of the specimen imaged by correlative electron microscopy (Diercks et al., 2013). These can introduce structural damage from electron bombardment and lower the yield of atom probe measurements (Herbig & Kumar, 2021). In contrast, the coating acts as a clear marker to identify the original specimen, including its native oxide layer.

It was also reported that the lensing system used to increase the field-of-view in the Invizo instrument leads to severe distortions in the reconstruction (Tegg et al., 2023). These are related to strong compression of the field lines and trajectories that are not appropriately taken into consideration in the currently available algorithms based on a point-projection model, making accurate reconstruction difficult. Cr coating does not lead to a significant change in the compression, meaning that current reconstruction algorithms remain applicable, and no noticeable change or distortions of the spherical δ'-precipitates can be observed in Figure 11.

### 4.6.2 Limitation

The coating is sometimes imaged only on one side, as in Figure 9B. This could be related to a slight tilt of the specimen's main axis with respect to the microscope's optical axis, or to a misalignment of the specimen with respected to the local electrode or the detector. In this case, the estimated ICF appears to be higher by 6% between the (002) and (113) pole marked in the figure, compared to between the (002) and the two (113) poles located on the perpendicular zone axis. Differences in the compression for specimen that do not exhibit cylindrical symmetry had previously been reported (Larson et al., 1999), which would lead to slight distortions in the reconstructed data since a single ICF is assumed across the field-of-view. This does not appear to be related to the location where the laser hits the specimen (bottom left in the detector hit map). The presence of a thicker coating on one side could be due to a slight misalignment during the deposition process, or the specimen bending slightly, as was sometimes observed when the shank angle is too low and/or radius of the specimens are too small (see Figure S3 for Al). This explains the apparent change in the field-of-view between the datasets in Figure 9A and B is related to the fact that, after coating, the imaged cross section is only on one the side of the original specimen, as illustrated in the Figure S4.

From the detector hit maps of the coated specimen (Figure 9B), many areas of higher and lower event density appear, forming a speckled pattern indicative of overheating of the specimen from the laser. This observation suggests that the absorption of laser energy by Cr in the UV region is significantly higher than that of pure Al. Therefore, more energy is absorbed in the form of thermal energy which heats the Al sample significantly (Vurpillot et al., 2009). Several publications have reported that Cr exhibits photoexcitation, which contribute to the higher sample temperature (Al-Kuhaili & Durrani, 2007; Khalaf et al., 2021). The presence of the oxide layer around the Al, and the multiple interfaces can also limit the heat flows away from the apex. However, the higher absorption of the Cr layer can also be a great advantage for materials that have very poor absorption properties for the laser at the used wavelength (here, $\lambda$=355 nm), such as glass (Kellogg, 1982) or (bio)minerals (Reddy et al., 2020; Grandfield et al., 2022), or most metals at shorter wavelengths. This would allow significantly lower laser energies to be used, reducing the base voltage, and extending the lifetime of the samples, along with reducing thermal tailing.

In addition, there can also be severe interference in the mass spectrum between peaks from the materials being analyzed, the coated metal and its possible oxides - the APT analysis of Cr oxides shows many series of peaks up to relatively high mass-to-charge state ratios (La Fontaine et al., 2015). It is therefore important to select the coating material in such a way that the expected signals from both the analyzed material and the coating material do not severely overlap in the mass spectrum. The interference between signals from the coating and the material of interest is readily visible in Figure 8. To avoid misinterpretation, it is often simpler to extract the inner part of the dataset into a separate reconstruction in the analysis software by selecting the appropriate region of the voltage curve and detector region-of-interest or extracting from a Cr-isosurface for instance and reassigning the peaks in the mass spectrum. At this stage, it is unclear whether the sputtered metal reacts with the surface of the original specimen or its native oxide, or if the deposition process causes structural damage or implantation subsurface. This will be studied in future work.

### 4.6.3 Future considerations

Our coating approach is versatile, and many parameters can be optimized. For example, the coated metal can be chosen to match the evaporation field of the coated material to minimize the evaporation field contrast and possible resulting aberrations.

There are many possibilities for the metal to be coated, and it is easy to change the target, but the deposited metal must adhere sufficiently and not react with the incident ion beam. In this respect, the choice of an element with only one or two isotopes, with relatively well-defined oxides, or with a low tendency to form oxides, or with high atomic masses, may seem optimal. Noble metals such as Ag, Pt, Pd or Au could also avoid possible reactions with the surface.

For each material, suitable sputtering parameters must first be found and optimized in the FIB, but this process offers the possibility of using a wide range of metals. Ti, In,

Bi, Al, Co and Ag have also been tried out with similar deposition parameters (see Figure S5-S10).

The Ag layer was very inhomogeneous and will require future optimization. Nevertheless, the mass spectrum of the Ag coating layer shows promising behavior, with only Ga and two unidentified peaks appearing at lower mass-to-charge state ratios than Ag. At higher masses, several Ag oxide peaks appear, which may also be related to the speckled texture and its non-uniform evaporation behavior (Figure S6). For In, the reaction with the incident Ga ions from the primary beam resulted in an uneven deposition from a possibly liquid source, as the In-Ga alloy (42,6 (In) /52,4 (Ga) at. %) system (Anderson & Ansara, 1991) has a low temperature eutectic (see Figure S7). This could be optimized by using the Xe-PFIB.

Similar observations were made for Al coated on a biomineral, where the Ga content in the sputtered layer was approx. 2.17±0.33 at. %, but the deposited layer was very homogeneous and uniform. This effect could be avoided by using Xe PFIB, whereas Al is a promising candidate as a low evaporation material for coatings (Figure S5). Only two candidates that have so far shown good sputtering properties with the same parameters are Ti, which has five isotopes and is therefore not perfect choice (Figure S10), due to the high probability of overlap and the formation of oxides, which further complicate the mass spectrum. The second candidate is Co, which also forms a uniform layer and has only one isotope, which make Co attractive for a coating material, because the probability for an overlap is lower. There remains a possibility that Co overlaps with some peaks of interest, and a single isotope may prevent decomposition and improvement in the composition measurement. However, Co has a higher evaporation field than all the other materials used in these studies, which needs to be considered when choosing the coating material (Figure S9).

The energy distribution of the deposited atoms or secondary ions emitted by the target is primarily undetermined at this stage, and so is how these ions possibly affect the specimen's surface upon deposition – i.e. causing implantation or structural damage to the surface and subsurface layers. Adjusting the energy of the primary Ga or Xe beam may be a lever to adjust this, to minimize damage from these secondary particles.

Finally, previous studies have suggested that a coated layer can be used to increase the electrical conductivity of the specimen, allowing the measurement of insulators in voltage-pulsing mode (Adineh et al., 2016, 2017, 2018). An improved electrical conductivity of the samples can be seen in Figure S5. Prior to the coating, a clear charging effect can be observed due to the poor electrical conductivity of the specimen in the electron beam, which disappears after the Cr coating, indicating charge transport to the shank of the specimen. The effect of these coatings on electrical and thermal conductivity provides the motivation to carry out simulations in the future to understand their effect.

Here, we only used laser-pulsed atom probe, which has more parameters to be adjusted (wavelength, laser energy, frequency) to optimize the chemical composition (Mancini et al., 2014). In addition, the increase in temperature can lead to diffusion processes on the surface, which have a negative effect on the accuracy of the reconstruction and thus lead to artefacts (Larson et al., 2013; Lefebvre-Ulrikson et al., 2016). On the opposite side of the laser incident, a "shadow effect" occurs. This reduces the mass resolution of the APT, especially for materials with poor thermal conductivity (Sha et al., 2008). Coating poorly conducting materials with a metallic layer and measuring in voltage mode would be an interesting approach to avoid the negative effects of using thermally induced field evaporation. The field evaporation contrast may lead to a build-up of stresses at the interface between the specimen and the coating and facilitate fracture, or simply create additional aberrations, which motivates further studies.

## 5 Summary & conclusions

To summarize, we have demonstrated a low-cost and fast method to form a conformal coating on APT specimens that appears homogeneous and uniform. The method uses *in-situ* sputtering of a metallic target by the FIB. This approach is versatile and leads to many benefits including:

(i) environmental protection during transfer between the FIB and the atom probe, which is critical for studies of e.g. surfaces and reactive materials;

(ii) a higher sample yield, possibly due to the mechanical stabilization provided by the coating;

(iii) improved mass resolution and background level leading to improved sensitivity by a better thermal conductivity of and or better absorption of the laser energy by the metallic coating;

(iv) possibilities to image the shank of the specimens, which dramatically increases the surface area of e.g. native oxides formed on atom probe specimens and further stabilization of the oxide layer evaporation;

(v) the possibility to detect larger numbers of precipitates due to a larger field of view within the same specimen, thus increasing the throughput and statistical power and decreasing the number of runs and lower the cost;

(vi) increasing the field of view by increasing the radius of the original tip does not compress the field lines, so the existing reconstruction algorithm is retained.

Much remains to be explored, such as optimizing the sputtering parameters and their influence on the thickness or crystalline structure of the coating, adapting the sputtering parameters to different target materials, the influence of the coating on non-conductive or poorly thermally conductive materials, or the influence of the electrical and thermal conductivity of the target metal on analytical performance.

# Declaration of Competing Interest

The authors declare that they have no known competing financial interest or personal relationships that could have appeared to influence the work reported in this paper.

# Acknowledgments

Uwe Tezins, Andreas Sturm and Christian Broß are acknowledged for their support to the FIB & APT facilities at MPIE. TMS, IM, EVW, BG are grateful for funding from the DFG through the award of the Leibniz Prize 2020. IM, EVW, BG are grateful for funding from the ERC – SHINE (771602). MK and BG acknowledge financial support from the German Research Foundation (DFG) through DIP Project No. 450800666. XC is funded from DFG project ID 405553726-TRR 270, project Z01. CJ is grateful for financial support from Alexander von Humboldt Foundation. KJ is grateful for funding from the Samsung Electro-Mechanics. SHK acknowledge the KIAT grant funded by the Korea Government MOTIE (P0023676).# 6 References

ADINEH, V. R., MARCEAU, R. K. W., CHEN, Y., SI, K. J., VELKOV, T., CHENG, W., LI, J. & FU, J. (2017). Pulsed-voltage atom probe tomography of low conductivity and insulator materials by application of ultrathin metallic coating on nanoscale specimen geometry. *Ultramicroscopy* **181**, 150–159.

ADINEH, V. R., MARCEAU, R. K. W., VELKOV, T., LI, J. & FU, J. (2016). Near-Atomic Three-Dimensional Mapping for Site-Specific Chemistry of 'Superbugs'. *Nano Letters* **16**, 7113–7120.

ADINEH, V. R., ZHENG, C., ZHANG, Q., MARCEAU, R. K. W., LIU, B., CHEN, Y., SI, K. J., WEYLAND, M., VELKOV, T., CHENG, W., LI, J. & FU, J. (2018). Graphene-Enhanced 3D Chemical Mapping of Biological Specimens at Near-Atomic Resolution. *Advanced Functional Materials* **28**, 1801439.

AL-KUHAILI, M. F. & DURRANI, S. M. A. (2007). Optical properties of chromium oxide thin films deposited by electron-beam evaporation. *Optical Materials* **29**, 709–713.

ANDERSON, T. J. & ANSARA, I. (1991). The Ga-In (Gallium-Indium) System. *Journal of Phase Equilibria* **12**, 64–72.

ANWAY, A. R. (2003). Field Ionization of Water. *The Journal of Chemical Physics* **50**, 2012–2021.

BALOGH, Z., REDA CHELLALI, M., GREIWE, G.-H., SCHMITZ, G. & ERDÉLYI, Z. (2011). Interface sharpening in miscible Ni/Cu multilayers studied by atom probe tomography. *Applied Physics Letters* **99**, 181902.

BARROO, C., AKEY, A. J. & BELL, D. C. (2020). Aggregated nanoparticles: Sample preparation and analysis by atom probe tomography. *Ultramicroscopy* **218**.

BOSTEL, A. & YAVOR, M. (2010). (12) Patent Application Publication (10) Pub. No.: US 2010/0223698 A1. **1**.

BUNTON, J. H., OLSON, J. D., LENZ, D. R. & KELLY, T. F. (2007). Advances in pulsed-laser atom probe: instrument and specimen design for optimum performance. *Microscopy and microanalysis : the official journal of Microscopy Society of America, Microbeam Analysis Society, Microscopical Society of Canada* **13**, 418–427.

CEREZO, A., WARREN, P. J. & SMITH, G. D. W. (1999). Some aspects of image projection in the field-ion microscope. *Ultramicroscopy* **79**, 251–257.

DE GEUSER, F. & GAULT, B. (2017). Reflections on the Projection of Ions in Atom Probe Tomography. *Microscopy and Microanalysis* **23**, 238–246.
29

# Supporting Information

Tabel S1: Parameters of the annular pattern at 30Kv, 40pA for the deposition of Cr on a Thermofisher FIB.

| Outer Diameter | 12 µm |
|---|---|
| Inner Diameter | 8 µm |
| Z-size | 1 µm |
| Scan direction | Inner to Outer |
| Dewll time | 1.00 µs |
| Overlap X | 50% |
| Overlap Y | 50% |
| Scan type | Circular |
| Fill style | Solid |
| Dose | $2.70\text{E}^{-1}$ µm$^3$/nC |
| Total Volume Sputter Rate | $1.08\text{E}^{-2}$ µm$^3$/s |

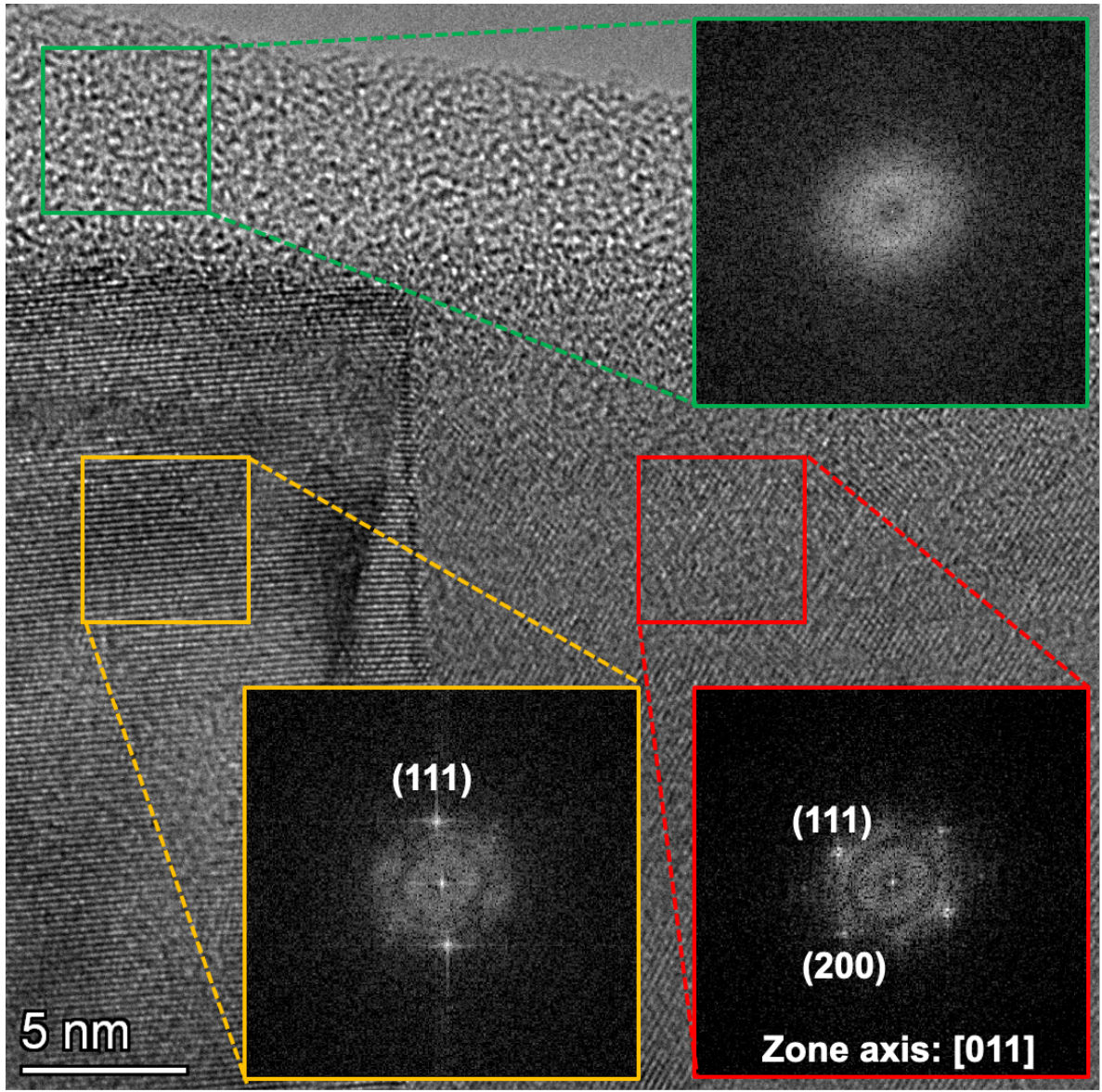

Figure S 1: fast-Fourier transform (FFT) obtained in the red and yellow boxes within the micrograph from the two Ni-grains imaged in Figure 4D, showing spots corresponding to the crystal of Ni. The FFT for the Cr-coating, obtained from the green box, displays only rings associated to an amorphous structure.

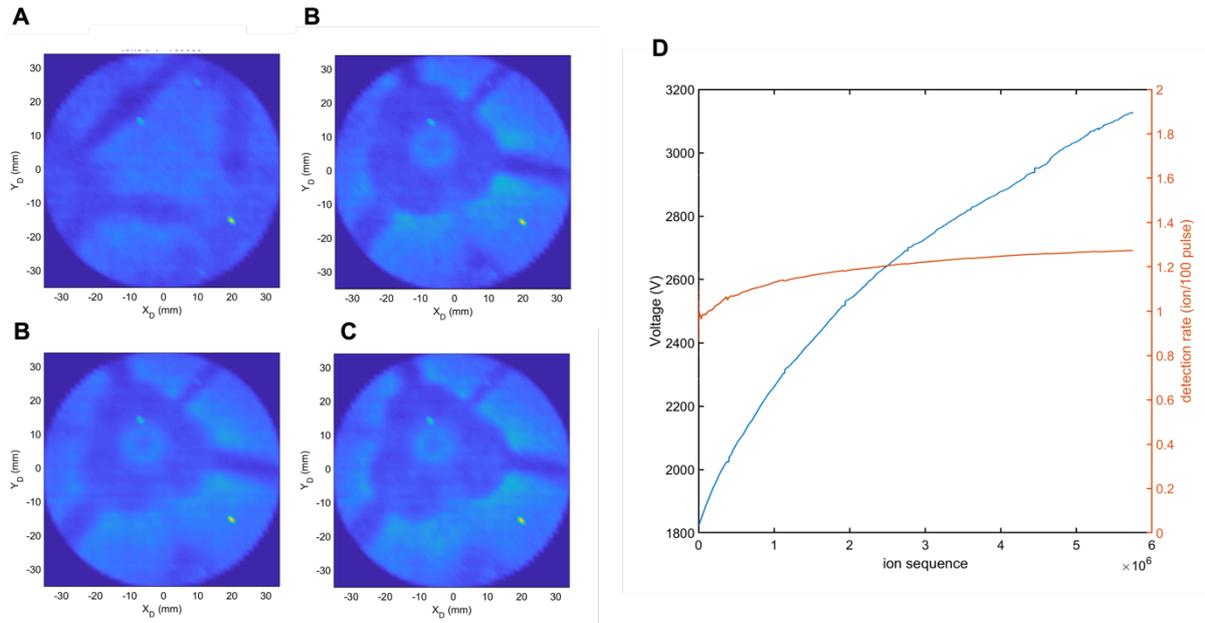

Figure S 2: (A-C) shows different detector hit maps during the measurement, indicating the different crystallographic orientations of the sample. In (D) the voltage and detection curves are plotted, there is no correlation between the voltage dips and an increased evaporation rate due to microfractures.

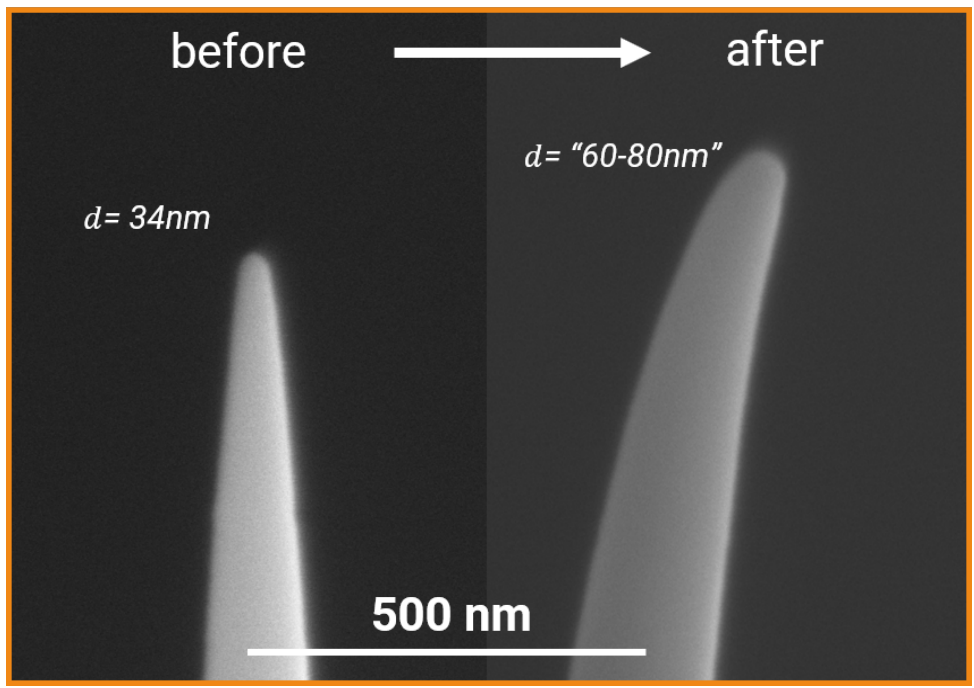

Figure S 3: Al sample before and after Cr-coating. The coating is uniform, and the radius of the original tip has increased but the tip is slightly bent.

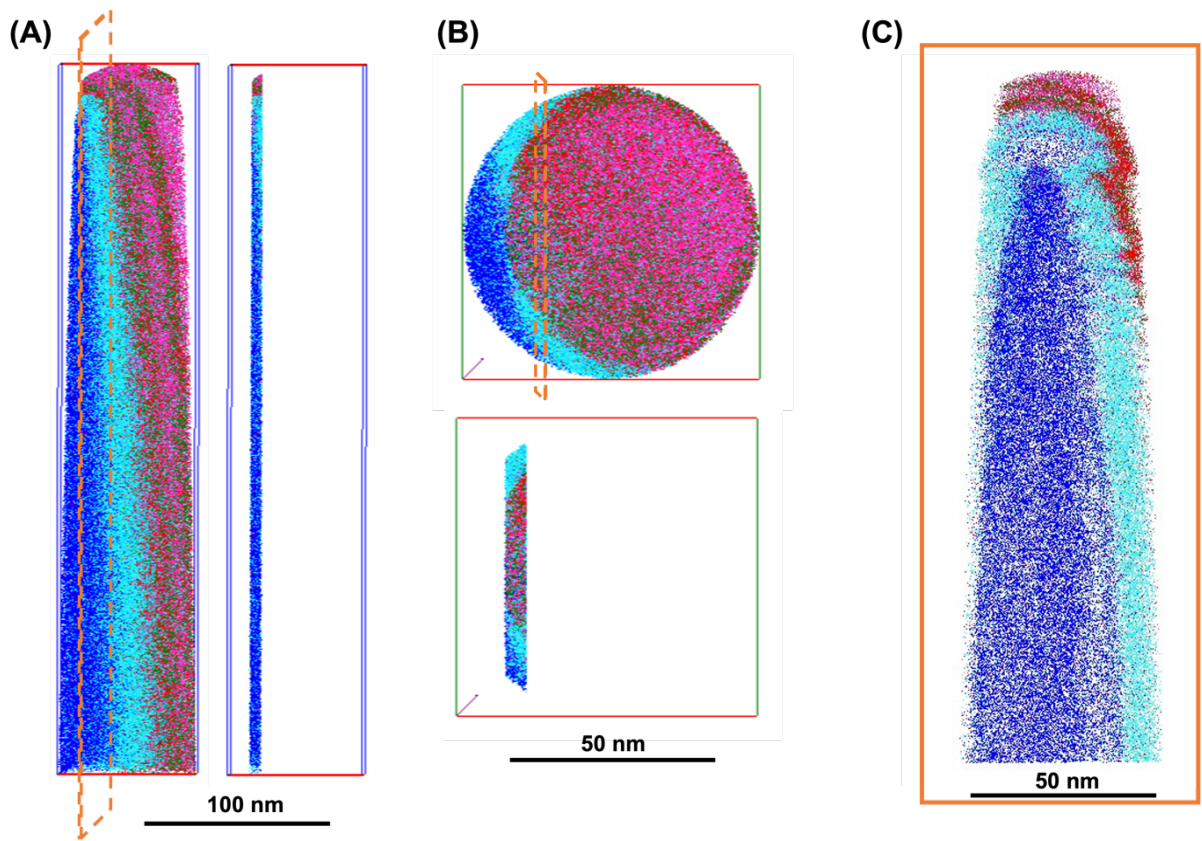

Figure S 4: Full view of the dataset obtained for an Al-specimen coated with Cr, reported in Fig. 9.

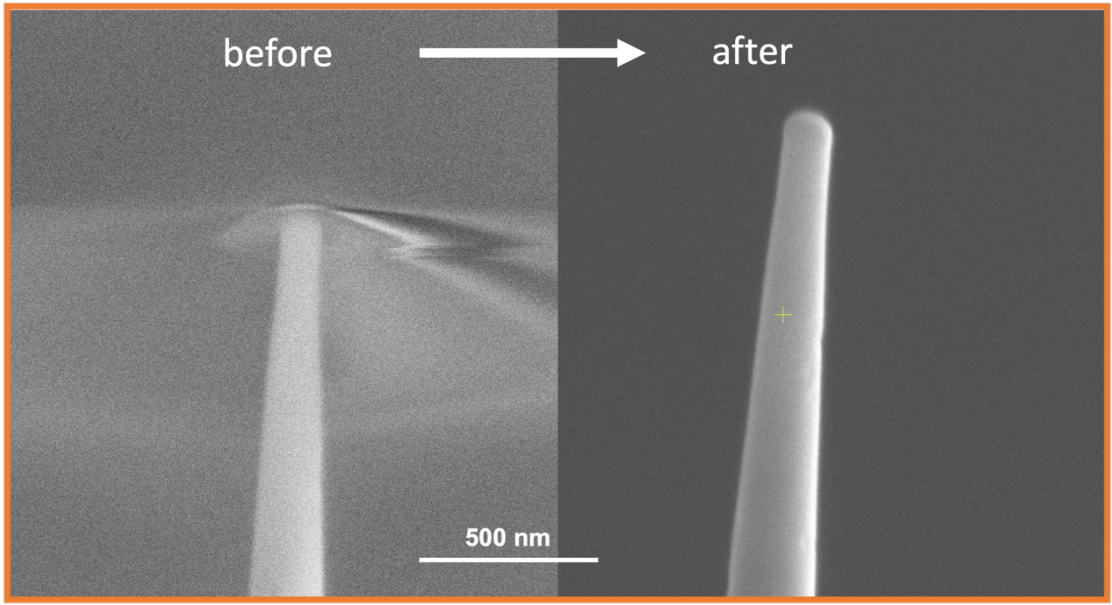
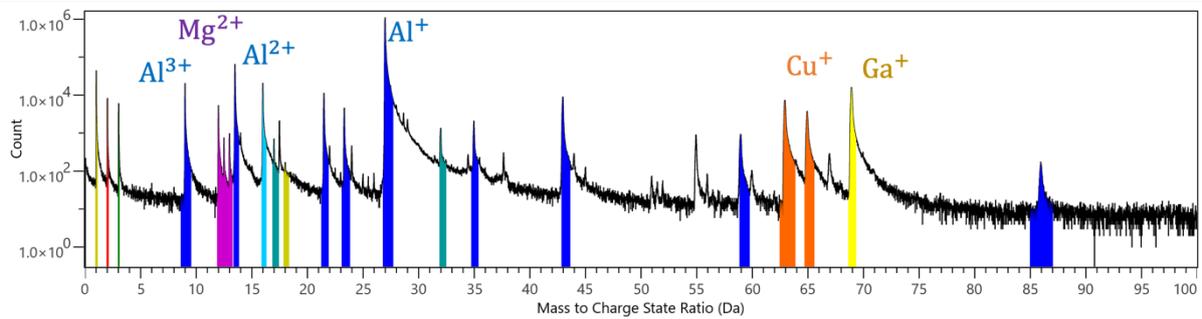

Figure S 5: Specimen coated with Al. The coating is uniform, the extracted mass spectra of the Al coating show some impurities (Mg and Cu) of the sputtered Al lamellae and several peaks of Al oxides.

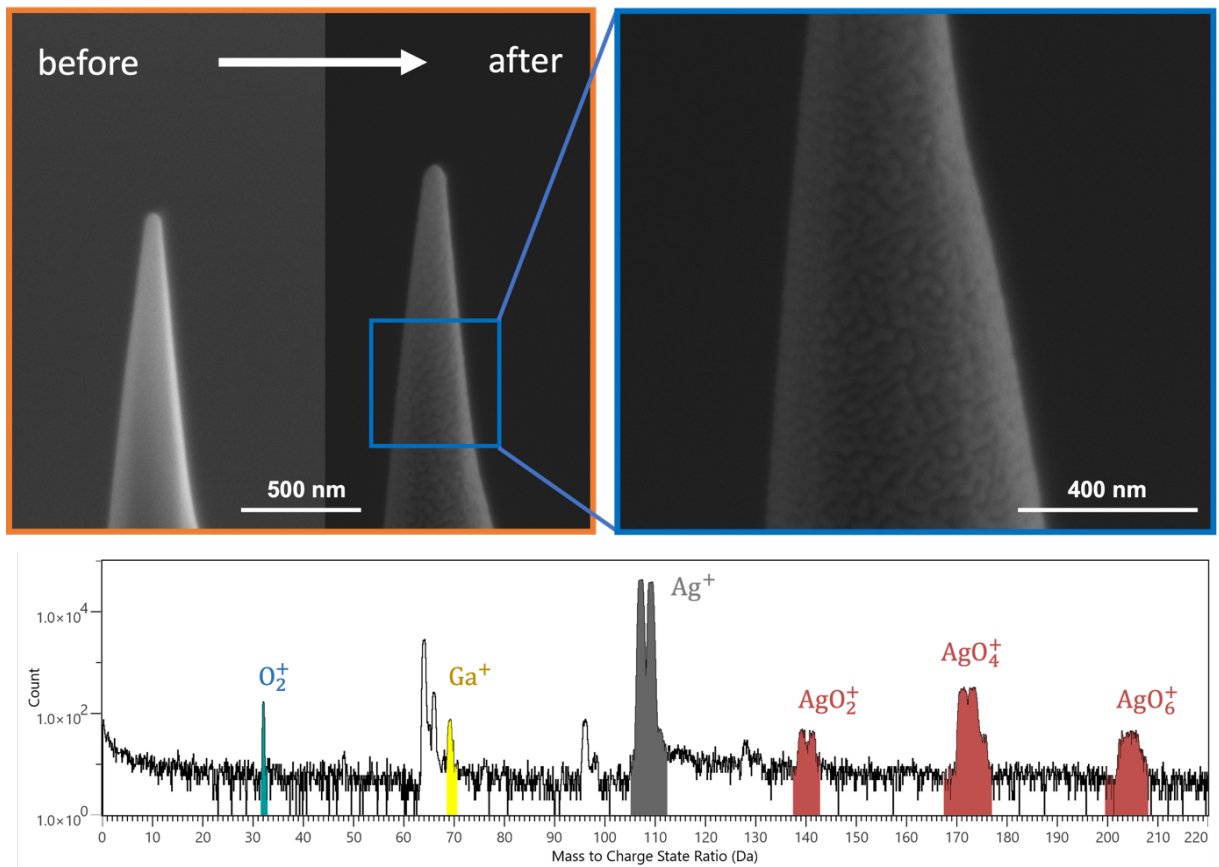

Figure S 6: Al specimen coated with Ag. The coating is not uniform and has a speckled texture. The mass spectrum of the Ag coating shows promising behavior as at low mass to charge state ratios only Ag-Ga and two unidentified peaks appear. At higher masses, several Ag oxide peaks appear, which may also be related to the speckled texture and its non-uniform evaporation behavior.

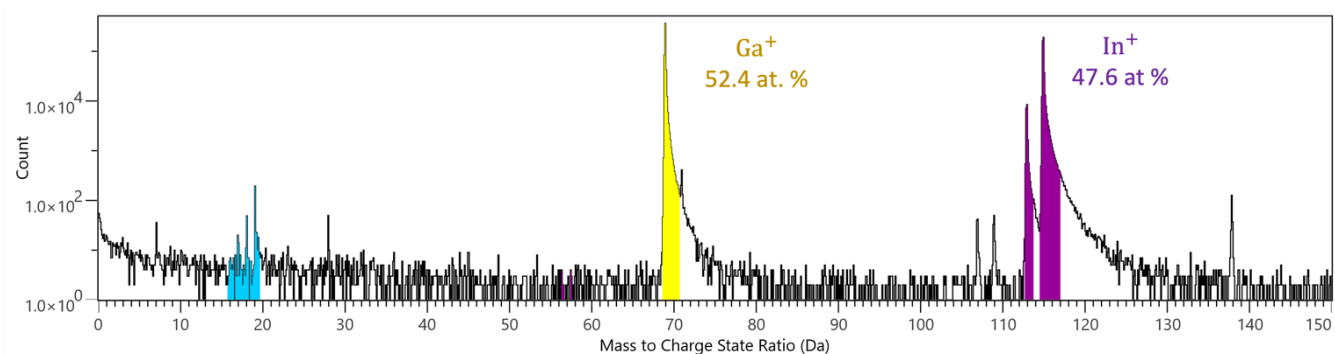

Figure S 7: Mass spectrum of the coated In layer, reveals a high proportion of Ga.

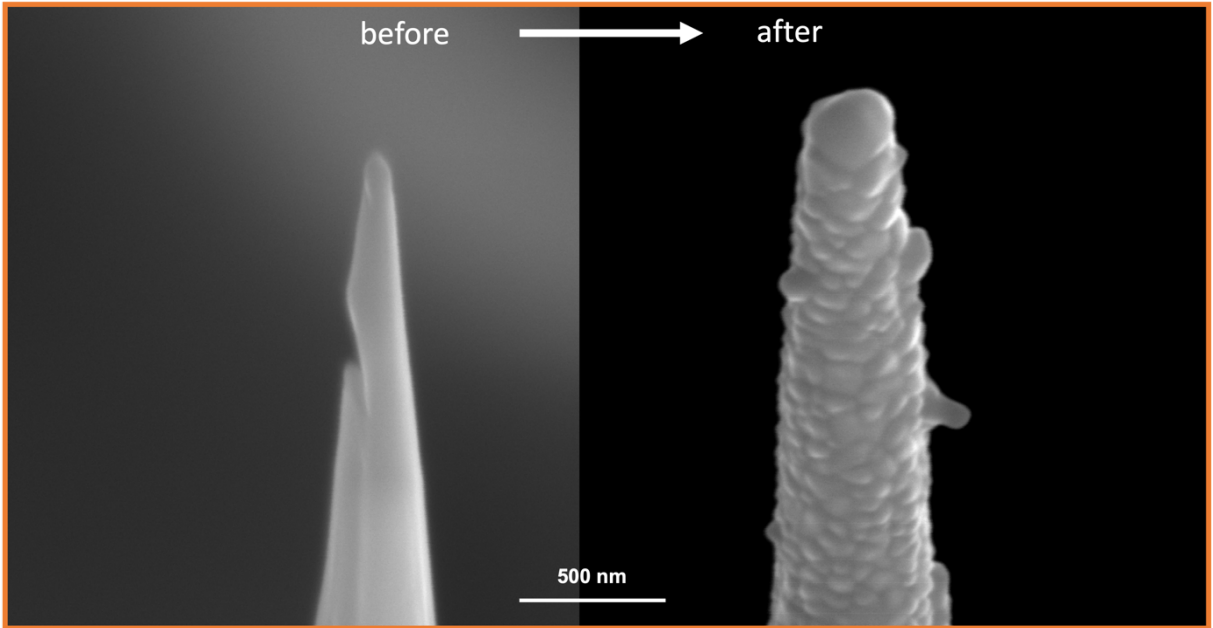

Figure S 8: Bismuth coating exhibits a very inhomogeneous coating and a very strong sputtering rate.

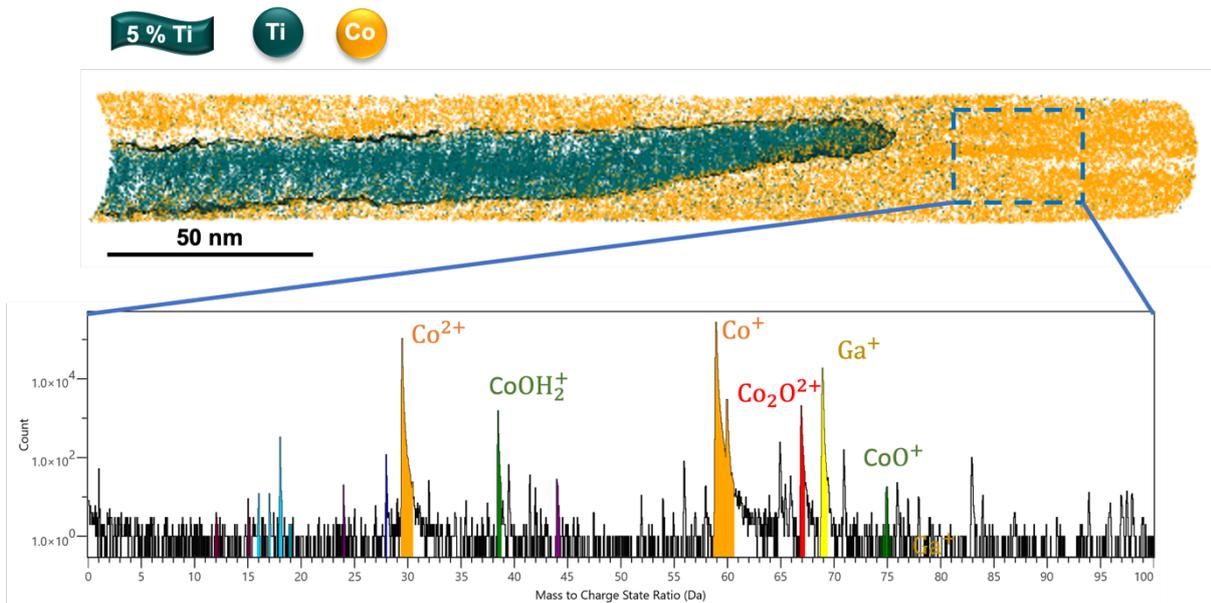

Figure S 9: Co coating shows a homogeneous coating around the specimen, the mass spectrum from the cobalt layer reveals that Co barely forms any oxides.

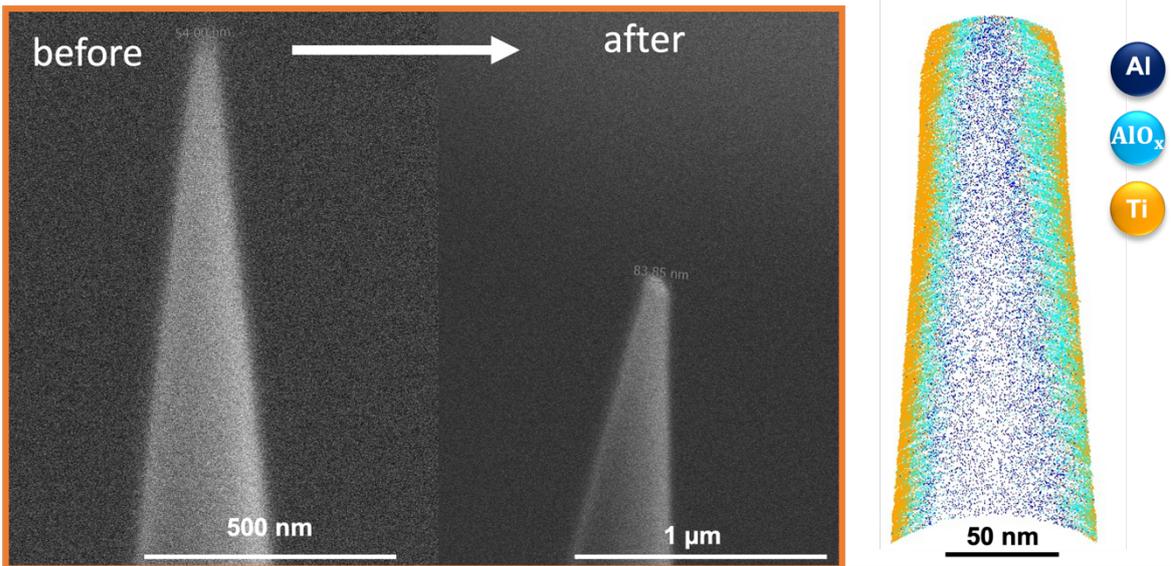
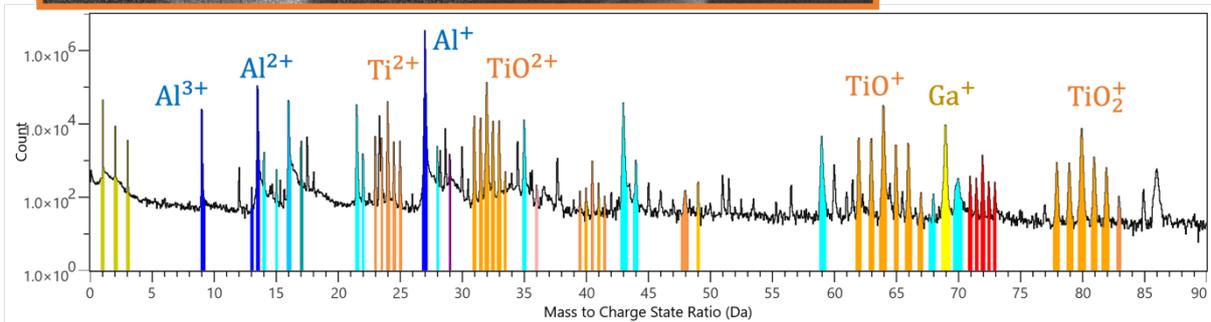

Figure S 10: Al specimen coated with Ti, indicates a uniform coating. However, the mass spectrum reveals that Ti has a very complex mass spectrum due to its isotopes and tendency to form oxides.